\def\Cim{Distinguished rheological models for solids in the framework of a
thermodynamical internal variable theory}

\documentclass[10pt,a4paper]{article}

\newcount\relabeling\relabeling=0  
\newcount\pageeqno\pageeqno=0  
\newcount\figures\figures=0  
\newcount\preview\preview=0    

\ifnum\preview=0  
\ifnum\preview=1  
\ifnum\preview=2  
  \AtBeginDocument{\Large}\makeatletter
  \def\section{\@startsection {section}{1}{\z@}{-3.5ex\@plus-1ex
    \@minus-.2ex}{2.3ex\@plus.2ex}{\normalfont\LARGE\bfseries}}
  \def\subsection{\@startsection{subsection}{2}{\z@}{-3.25ex\@plus-1ex
    \@minus-.2ex}{1.5ex\@plus.2ex}{\normalfont\Large\bfseries}}
  \makeatother\fi

\newcommand\nc{\newcommand*}  \nc\longnc{\newcommand}

\usepackage{amsmath}  

\long\def\OMIT#1{\relax}       

                    \let\Z\kern  
         
  \def\YY#1{\setbox1\HB to0em{\ }\RB{#1}}
     \def\yy#1{\YY{#1ex}}
          
\let\HB\hbox  \def\SB{\setbox1\HB}  \def\CB{\copy1}    
                \def\CC{\copy2}    
\def\RB#1{\raise#1\CB}    \def\YB{\ht1}  
\def\RC#1{\raise#1\CC}      
\newcount\n  \newdimen\w  \newdimen\h   
\def\mathsizes#1#2#3{\mathchoice{#1}{#1}{#2}{#3}}   

\def\tS#1{\ifcase#1\tiny\or          
  \scriptsize\or\footnotesize\or\small\or\normalsize\or\large\or
  \Large\or\LARGE\or\huge\or\Huge\else\ifnum#1<0\tiny\else\Huge\fi\fi}
\makeatletter                        
\def\cS{\ifx\@currsize\normalsize    
  4\else\ifx\@currsize\small 3\else\ifx\@currsize\footnotesize
  2\else\ifx\@currsize\large 5\else\ifx\@currsize\Large
  6\else\ifx\@currsize\LARGE 7\else\ifx\@currsize\scriptsize
  1\else\ifx\@currsize\tiny  0\else\ifx\@currsize\huge
  8\else\ifx\@currsize\Huge  9\else 4\fi\fi\fi\fi\fi\fi\fi\fi\fi\fi}
\DeclareRobustCommand\rS[1]{\ifmmode\@nomath\rS\else    
  \@tempcnta\cS\advance\@tempcnta#1\relax\tS\@tempcnta} 
\makeatother
\def\mS#1{\ifcase#1\displaystyle\or  
  \textstyle\or\scriptstyle\or\scriptscriptstyle\else\textstyle\fi}
\def\dS#1{\csname\ifcase#1relax\or   
  relax\or big\or Big\or bigg\or Bigg\fi\endcsname}

\def\textinmath#1{{\mathsizes                
  {\HB{#1}}{\HB{\tS1#1}}{\HB{\tS0#1}}}}      
\def\txt#1#2#3{\mskip#1mu                  
  \textinmath{#3}\mskip#2mu\relax }        
\let\mathcl\mathcal                      

\def\re#1{(\ref{#1})}   

\def\sect#1#2{{\section{#2}\label{#1}}}

\def\ssect#1#2{\subsection{#2}\label{#1}}

\long\def\quot#1{`#1'}    

\let\emp\textit       
\let\lat\textit                 
\def\ie{\lat{i.e.,\ }}  \def\etal{\lat{et al.\ }}    
\def\eg{\lat{e.g.,\ }}  \def\insitu{\lat{in situ}}  
  
\def\m#1{\scase=0$      #1         $}  
\def\mm#1{\m{       \,  #1  \,     }}  

\ifnum\pageeqno=1  
\usepackage{refcount}
\newcounter{lasteqpage}
\newcounter{lastlasteqpage}
\newcounter{lasteqno}
\newcounter{lastlasteqno}
\def\elabel#1{\label{#1}
  \setcounter{lastlasteqpage}{\thelasteqpage}
  \setcounter{lastlasteqno}{\thelasteqno}
  \setcounter{lasteqpage}{\getpagerefnumber{#1}}  
  \ifnum\thelasteqpage>\thepage\else\setcounter{lasteqpage}{\thepage}\fi
  \ifnum\thelasteqpage>\thelastlasteqpage\setcounter{lasteqno}{1}
  \else\addtocounter{lasteqno}{1}\fi} 
\def\non{\tagg\setcounter{lasteqpage}{\thelastlasteqpage}
              \setcounter{lasteqno}{\thelastlasteqno}}

\else  \def\elabel#1{\label{#1}}  \def\non{\tagg}  \fi

\makeatletter\@ifpackageloaded{amsmath}{
\def\eq#1#2{ \scase=1 \begin{align} \elabel{#1} #2 \end{align}}
\def\eqa{\eq}    
\def\eqn#1#2{ \scase=1 \begin{align} \elabel{#1} \non #2 \end{align}}
\def\eqan{\eqn}  
\def\lel#1{ \\ \elabel{#1}     }  
\def\leln#1{\\ \elabel{#1} \non}  
\def\tagg{\tag*{}}  
\def\mat#1{\begin{matrix} #1 \end{matrix}}
\def\smat#1{\begin{smallmatrix} #1 \end{smallmatrix}}
  }{
\def\eq#1#2{  \scase=0\begin{equation}    \elabel{#1}   #2 \end{equation}}
\def\eqa#1#2{ \scase=2\begin{eqnarray}    \elabel{#1}   #2 \end{eqnarray}}
\def\eqn#1#2{ \scase=0\begin{displaymath} \elabel{#1}   #2 \end{displaymath}}
\def\eqan#1#2{\scase=2\begin{eqnarray} \elabel{#1} \non #2 \end{eqnarray}}
\def\lel#1{\ifnum\scase=2\else\erroreqaneeded\fi \\ \elabel{#1}}
\def\leln#1{\ifnum\scase=2\else\erroreqaneeded\fi \\ \elabel{#1} \non}
\def\tagg{\nonumber}  
\def\lvert{|}  \def\rvert{|}  \def\lVert{\|}  \def\rVert{\|}
\def\mat#1{{\def\\{\cr}\matrix{#1}}}    
\def\smat#1{\hbox{\scriptsize{$\mat{#1}$}}}  
}\makeatother  

\newcount\scase  \def\7{&}  \def\s#1{\ifcase\scase#1\or\7#1\or\7#1\7\fi}

\def\0#1#2{\ifcase#1{#2}\or\lt(#2\rt)\or\lt[{#2}\rt]\or\lt\{{#2}\rt\}\or
  \mathord<{#2}\mathord>\or\lt\langle{#2}\rt\rangle\or\lt\lvert{#2}\rt
  \rvert\or\lt\lVert{#2}\rt\rVert\fi}
\def\1#1#2{\ifcase#1{#2}\or(#2)\or[#2]\or\{#2\}\or\mathord<{#2}\mathord
  >\or\langle{#2}\rangle\or\lvert{#2}\rvert\or\lVert{#2}\rVert\fi}
\def\2#1#2{\ifcase#1{#2}\or\big(#2\big)\or\big[#2\big]\or\big
  \{#2\big\}\or\big<#2\big>\or\big\langle#2\big\rangle\or\big
  \lvert#2\big\rvert\or\big\lVert#2\big\rVert\fi}
\def\3#1#2{\ifcase#1{#2}\or\Big(#2\Big)\or\Big[#2\Big]\or\Big\{#2\Big
  \}\or\Big<#2\Big>\or\Big\langle#2\Big\rangle\or\Big\lvert#2\Big
  \rvert\or\Big\lVert#2\Big\rVert\fi}
\def\4#1#2{\ifcase#1{#2}\or\bigg(#2\bigg)\or\bigg[#2\bigg]\or\bigg
  \{#2\bigg\}\or\bigg<#2\bigg>\or\bigg\langle#2\bigg\rangle\or\bigg
  \lvert#2\bigg\rvert\or\bigg\lVert#2\bigg\rVert\fi}
\def\5#1#2{\ifcase#1{#2}\or\Bigg(#2\Bigg)\or\Bigg[#2\Bigg]\or\Bigg
  \{#2\Bigg\}\or\Bigg<#2\Bigg>\or\Bigg\langle#2\Bigg\rangle\or\Bigg  
  \lvert#2\Bigg\rvert\or\Bigg\lVert#2\Bigg\rVert\fi}
\def\9#1#2{\ifcase#1{#2}\or\left(#2\right)\or\left[#2\right]\or\left
  \{#2\right\}\or\left\langle{#2}\right\rangle\or\left\langle{#2}\right
  \rangle\or\left\lvert{#2}\right\rvert\or\left\lVert{#2}\right\rVert\fi}
\def\lt{\mathopen{}\mathclose\bgroup\left} \def\rt{\aftergroup\egroup\right}

\def\bit{    \mskip1mu}  \def\biT{    \mskip-1mu}      
\def\bitt{   \mskip2mu}  \def\biTT{   \mskip-2mu}      

\let\f\frac                        
\def\F#1#2#3#4#5{\1#3{\1#1{#4}/\1#2{#5}}} 
\def\largerfrac#1#2#3{\mathchoice  
  {\SB{$\mS0\vcenter{}$}\w=\YB\SB{\rS{#1}$\mS0\vcenter{}$}
    \advance\w by-\YB\raise\w\HB{\rS{#1}$\mS0\frac{#2}{#3}$}}
  {\SB{$    \vcenter{}$}\w=\YB\SB{\rS{#1}$    \vcenter{}$}
    \advance\w by-\YB\raise\w\HB{\rS{#1}$    \frac{#2}{#3}$}}
  {\SB{$\mS2\vcenter{}$}\w=\YB\SB{\rS{#1}$\mS2\vcenter{}$}
    \advance\w by-\YB\raise\w\HB{\rS{#1}$\mS2\frac{#2}{#3}$}}
  {\SB{$\mS3\vcenter{}$}\w=\YB\SB{\rS{#1}$\mS3\vcenter{}$}
    \advance\w by-\YB\raise\w\HB{\rS{#1}$\mS3\frac{#2}{#3}$}}}

\def\restr#1#2{{\lt.#1\rt|}_{#2}}  
\def\tr{\mathop{\txt00{tr}}}

\def\e{\mathrm{e}}  
\def\dd{\mathrm{d}}    
\let\pd\partial

\def\tensUpRt#1{^{\mathrm{#1}}}            
\def\symm{\tensUpRt{S}}   \def\asymm{\tensUpRt{A}}
\def\dev{\tensUpRt{d}}   \def\sph{\tensUpRt{s}}
\let\tens\mathbf  
\let\Tens\mathBf  

\def\Ded{\mathcl{E}\dev}  \def\Des{\mathcl{E}\sph}
\def\Dsd{\mathcl{S}\dev}  \def\Dss{\mathcl{S}\sph}
\def\parconn{\,||\,}  \def\serconn{\biTT\sim\biTT}  
\def\paraline{\rule{.03em}{.9ex}}
\def\para{_{}^{\paraline\kern.06em\paraline}}  
\def\orth{_{}^{\hbox{\large$\mS3\perp$}}}      
\def\model#1#2{\hbox{($#1 \asymp #2$)}}

\def\qkv{Kluitenberg--Verh\'as}

\def\qel#1{{#1}_{\txt00{el\yy{1.9 }}}}
\def\qanel#1{{#1}_{\txt00{anel\yy{1.9 }}}}
\def\qirr#1{{\hat{#1}}}
\def\qth#1{{#1}_{\txt00{th\yy{1.9 }}}}
\def\qcurr#1#2{{#1}_{#2}}  
\def\qini{}
\def\qext#1{{\tilde{#1}}}
\def\qindb{I_1}  \def\qindc{I_2}

\def\qa{a}  
\def\qb{b}  
\def\qc{c}  
\def\qe{e}
\def\qj{j}  \def\qqj{\tens{\qj}}
  
\def\ql{l}
\def\qm{m}
\def\qs{s}
\def\qD{\f{\dd}{\dd t}}  \def\qDD{\f{\dd^2}{\dd t^2 }}
\def\qqD{D}  \def\qqS{S}
\def\qT{T}

\def\qyoung{E}
\def\qalp{E_0}
\def\qbet{E_1}  \def\qgam{E_2}  \def\qtau{\tau}
\def\qeps{\varepsilon}
\def\qqeps{\Tens\qeps}
\def\qlam{\lambda}
\def\qrho{\varrho}
\def\qsig{\sigma}  \def\qqsig{\Tens{\qsig}}
\def\qxi{\xi}  \def\qqxi{\Tens{\qxi}}
\def\qeta{\zeta}  
\def\qentprod#1{\pi_{\biT #1}^{}}

\begin{document}  

\title{\Cim} 
 \author{
Csaba Asszonyi$^{1}$, Tam\'as F\"ul\"op$^{1, 2}$ and Peter V\'an$^{1, 2,
3}$
 \\
{\footnotesize
$^{1}$ Montavid Research Group for Thermodynamics, Igm\'andi u. 26,
H-1112 Budapest, Hungary}
 \\
{\footnotesize
$^{2}$ Dept.\ of Energy Engineering, Budapest Univ.\ of Technology and
Econ., PO Box 91, H-1521 Budapest, Hungary}
 \\
{\footnotesize
$^{3}$ Wigner Research Centre for Physics, PO Box 49, H-1525 Budapest,
Hungary}
 }

\date{
}  
\maketitle

 \begin{abstract}
We present and analyze a thermodynamical theory of rheology with single
internal variable. The universality of the model is ensured as long as
the mesoscopic and/or microscopic background processes satisfy the
applied thermodynamical principles, which are the second law, the basic
balances and the existence of an additional---tensorial---state
variable. The resulting model, which we suggest to call the
\qkv\ body, is the Poynting--Thomson--Zener body with an
additional inertial element, or, in other words, is the extension of
Jeffreys model to solids. We argue that this \qkv\ body is the
natural thermodynamical building block of rheology. An important feature
of the presented methodology is that nontrivial inequality-type
restrictions arise for the four parameters of the model. We compare
these conditions and other aspects to those of other known
thermodynamical approaches, like Extended Irreversible Thermodynamics or
the original theory of Kluitenberg.
 \end{abstract}

\sect{.1..1.}{Introduction}


The method of internal variables can be considered as a universal
modelling tool for classical, macroscopic field theories. It is
universal in the sense that one introduces minimal assumption about the
physical mechanism of the modelled phenomena. An additional field
variable is the starting point. Its relation and coupling to existing
physical quantities and its evolution equation are the key questions
that are to be answered. Irreversible thermodynamics proposes a
particular method in this respect \cite{MauMus94a1,MauMus94a2}. The
general idea is that one can obtain the form of the evolution equation
and also the connection to other processes considering only general
principles, first of all, the second law
\cite{ColGur67a,Kes93a1,Mus93a1}. Evolution equations derived from any
structural, mesoscopic or microscopic realization of the extra field
variable must belong to this general form, as long as they are
restricted by the same general principles. The first application of
thermodynamical ideas for continua is due to Eckart, who also
investigated deviations from ideal elastic behaviour, in another
wording, anelasticity \cite{Eck40a1,Eck48a}. A thermodynamical framework
for anelasticity with internal variables was first treated by Biot
\cite{Bio54a} and developed by Kluitenberg as a state variable theory
\cite{Klu62a1,Klu62a2,Klu62a3,Klu63a,Klu64a,Klu68a,KluCia78a,CiaKlu79a,Kui94b}.
Less systematic applications are popular for various phenomena in
solids, sometimes without any thermodynamical background
\cite{HorBam10a}.

The concept of internal variables has a long history
\cite{Bri43b,Mau99b}, and there are numerous different versions and
names. The name {\em internal degrees of freedom} was introduced in
thermodynamics as an extension of the configurational space and
originally denoted an extension of the deterministic field with
statistical and probabilistic aspects \cite{PriMaz53a,GroMaz62b}. This
is the origin of the so-called mesoscopic theories
\cite{BleAta91a,RegEta05a,Pap09a}. This meaning is strictly
distinguished from the macroscopic fields named under the same
terminology {\em internal degrees of freedom}, used by Maugin for field
theories with Lagrangian dynamics \cite{Mau90a,MauMus94a1}. The
so-called {\em internal variables of state} are again different, as
their evolution is relaxational and has a thermodynamic origin
\cite{ColGur67a,MauMus94a1}. In spite of the fine details where these
notions are different for different authors (controllability, boundary
conditions, weak nonlocality, etc.), we can find a sufficiently general
framework where these concepts coincide and appear as a powerful
modelling tool of modern continuum physics \cite{VanAta08a,VanEta14a}.
In what follows, we do not address these aspects but demonstrate the
constructive modelling power of a seemingly restricted conceptual
framework.

A specific version of the concept of internal variable is called
{\em dynamical degree of freedom} \cite{Ver97b,CiaVer93a}, which variable
becomes zero in local thermodynamical equilibrium, and thus quantifies,
along a process, the instantaneous deviation from equilibrium, so to
say, the amount of irreversibility present. This simple, natural and
general assumption, utilized by us here as well, is crucial. Other
approaches introduce various additional conditions due to the considered
direct interpretation of the state variables, hence, they lead to
different classification and different restrictions on the
constitutive coefficients.
For example, in Extended Irreversible Thermodynamics (EIT)
\cite{DauLeb90a,LebEta08b}, the additional state variable is the
dissipative stress, while in the original Kluitenberg theory the state
variable is anelastic strain, which additively modifies the elastic
strain of the rheological process.

The tensorial order of the internal variable can usually be deduced from
the properties of the modelled phenomenon. For example, for a simple
description of the damage of solids, a scalar variable may suffice,
characterizing the level of degradation of the structure of the solid.
For describing heat conduction effects beyond the Fourier theory, a
correction to the heat flux is expected, therefore, a vectorial internal
variable offers itself \cite{CiaVer90a,CiaVer91a,VanFul12a}. For the
description of rheological effects, corresponding to the fact that
stress and strain are symmetric tensorial quantities, the naturally
expected, and
here confirmed, internal variable is a symmetric second order tensor.

As shown below, for linear Onsagerian equations, one can eliminate this
internal state variable and obtains a linear relationship between stress,
strain and some of its time derivatives: the zeroth and first derivative
of stress and the zeroth, first and second derivative of strain. In
notation, we will denote this by \model{0, 1}{0, 1, 2}. Consequently,
this relationship covers a number of classic rheological models as
special subclasses: the Kelvin--Voigt model, which is the {\model{0}{0,
1}} case, the {\model{0, 1}{1}} Maxwell model, the {\model{0, 1}{0, 1}}
Poynting--Thomson--Zener model, and the {\model{0, 1}{1, 2}} Jeffreys
model. Therefore, the internal variable approach provides a universal
framework for discussing these models on a common ground, and, in
particular, to investigate the thermodynamical properties of these models.

More closely, we find here that one \model{0, 1}{0, 1, 2} model arises
for the relationship between the deviatoric part of stress and of strain
(and its derivatives), and another independent \model{0, 1}{0, 1, 2}
relationship emerges for the spherical (trace) part of these tensors.
This means four material coefficients for the deviatoric part and
another four for the spherical part. Remarkably, it is just this \m { 4
+ 4 = 8 } -parameter model that is found needed and satisfactory in the
Anelastic Strain Recovery method \cite{MatTak93a,Mats08a,LinEta10p}, an
experimental method determining underground 3D \insitu\ stress via
measuring the rheological relaxation of borehole rock samples, and which
has also been utilized recently for evaluating uniaxial experiments
stretching plastic samples \cite{Kocsis,Csatar}. In uniaxial loading
processes, the deviatoric and spherical parts get intertwined, and the
resulting relationship between stress and longitudinal strain turns out
to be a \model{0, 1, 2, 3}{0, 1, 2, 3, 4} model, as is shown in the
Appendix.

In a \model{0, 1}{0, 1, 2} model, the material coefficients cannot be
arbitrary because thermo\-dynamics---more closely, thermodynamical
stability and non-negative entropy production---reveals conditions on
them. A nontrivial finding below is that, even within the remaining
parameter region, only half of it can be represented via rheological
networks made of springs and dashpots. For the other half, an additional
element, the \model{0}{2} inertial element---introduced by Verh\'as
\cite{Ver85b,Ver97b}---, is also necessary.

There is an additional advantage of the elimination procedure: it makes
it manifest that the problem of boundary conditions does not emerge
here. This is actually due to the homogeneous ({\it i.e.,}
free-of-gradients) constitutive equations, and thus uniqueness of a
solution is ensured by the same boundary conditions that determine a
unique solution for the corresponding elasticity problem, only the
naturally expected further initial conditions have to be added
\cite{Ful09c}. In addition, in the light of a unified treatment with the
Maugin-type internal degrees of freedom and considering also a weakly
nonlocal extension, the question of boundary conditions can be handled
either by variational or direct thermodynamic prescriptions
\cite{VanAta08a,VanEta14a}.

Our treatment here is different from the ones given by Kluitenberg, by
EIT and by Verh\'as, and also from the approach of classical
irreversible thermodynamics, in an important respect. When introducing
the classic constitutive solution of the entropy inequality, assuming
linear relationships between the thermodynamical fluxes and forces, we do
not impose Onsager-Casimir symmetry relations. The principal reasons of
this choice is Occam's razor: without microscopic interpretation of the
internal variables, the conditions of Onsager need not apply
\cite{Ons31a1,Ons31a2}. This way the rightful criticism of Truesdell
\cite{Tru84b} does not hold for our treatment.
This generality has already proved to be crucial for the idea of dual
internal variables \cite{VanAta08a}, which turned out to be a powerful
unification method for waves in solids
\cite{BerEta10a1,Eng10a,EngBer12a,BerEng13a,EngBer13a}, and
also for deriving generalized mechanics \cite{BerEta11a, VanEta14a}.


In what follows, we develop the thermodynamical theory of rheology of
solids in the small strain approximation. First, we present the
essential steps in one spatial dimension, starting from the mechanical
properties and introducing the thermodynamical requirements. The
essential elements of the theory, as well as the most important
consequences, are manifest in this presentation. Then we develop the
three dimensional complete version. Finally, we discuss our approach and
compare it to the original assumptions and consequences by Kluitenberg
and by Extended Irreversible Thermodynamics. Two Appendices are devoted
to two special technical aspects: the combined behaviour of the
deviatoric and volumetric components in uniaxial loadings, and a variant
of the introduction of the internal state variable.

\sect{.4..2.}{The procedure in one spatial dimension}

\ssect{.4..2.1.}{The initial system}

According to the internal state variable methodology, first we have an
initial thermodynamical system, then we assume an additional internal
variable, shift the entropy by a concave expression of it, and ensure
the positive definiteness of entropy production via Onsagerian
equations. For our present purposes, the initial system is a linearly
elastic solid. To place the focus on the essential aspects, let us start
with the simple setting of the one space dimensional case.

Namely, we have a solid, with elastic strain (or deformedness, see
\cite{FulVan12a}) variable \m { \qqeps }, which is a scalar in one
dimension, and with elastic stress
 \eq{.4.1.}{
\qini\qsig \1 1 {\qeps} = \qyoung \qeps ,
 }
where, again for simplicity, the Young modulus \m { \qyoung } is treated
as constant. Assuming small deformations only, which is actually a
fairly good approximation for many situations concerning solids, the
velocity gradient is approximately equal to \m { \dot \qeps }, the time
derivative of strain, and the mechanical power exerted by \m { \qini\qsig
} is simplified to
 \eq{.4.2.}{
\qini\qsig \dot \qeps = \qyoung \qeps \dot \qeps =
\frac{\dd}{\dd t} \0 1 { \f {\qyoung}{2} \qeps^2 } =
\qrho \f {\dd}{\dd t} \0 1 { \f {\qyoung}{2\qrho} \qeps^2 } =
\qrho \qel{\dot{\qe}}
 }
with the {specific} elastic energy
 \eq{.4.3.}{
\qel{\qe} \1 1 {\qeps} = \f {\qyoung}{2 \qrho} \qeps^2
 }
and the mass density \m { \qrho }, which is constant within the range of
small deformations. Staying within the small-deformation regime, we also
do not have to distinguish among partial time derivative,
comoving/substantial time derivative and the various objective time
derivatives \1 1 {see \cite{Van08a1} for the finite deformation
differences among such derivatives}.

As the function of temperature \m { \qT } and strain \m { \qeps }, the
{specific} internal energy \m { \qini\qe } of the initial system is
considered as
 \eq{.4.4.}{
\qini\qe \1 1 {\qT, \qeps} = \qth{\qe} \1 1 {\qT} + \qel\qe \1 1 {\qeps} ,
 }
\m { \qth\qe \1 1 {\qT} } being related to the constant or nonconstant
specific heat \1 2 {\m { \qth\qe \1 1 {\qT} = \qc \qT } in the constant
case}. The separated variables in \re{.4.4.} indicate that thermal
expansion is now also disregarded.

Again if written as a function of temperature, the thermodynamically
corresponding {specific} entropy is \m { \qini\qs = \qini\qs \1 1 {\qT}
}, satisfying the thermodynamical consistency property
 \eq{.5.5.}{
\f {\dd \qini\qs}{\dd \qT} = \f {1}{\qT} \f{\dd \qth\qe}{\dd \qT} ,
 }
which follows---via \re{.4.4.}, \re{.4.3.} and \re{.4.1.}---from the
Gibbs relation \mm{ \qrho \dd \qini\qe = \qrho \qT \dd \qini\qs +
\qini\qsig \dd \qeps ,} rearrangable as
 \eq{.5.6.}{
\qrho \dd \qini\qs = \f{\qrho}{\qT} \dd \qini\qe -
\f{\qini\qsig}{\qT} \dd \qeps .
 }
For example, in the constant specific heat case,
 \eq{.5.7.}{
\qini\qs \1 1 {\qT} = \qc \ln \f { \qT}{\qT_0} + \qs_0
 }
is found, with auxiliary constants \m { \qT_0, \qs_0 }.

Taking \re{.4.2.} for the expression of power, the balance of internal
energy---\ie the first law of thermodynamics---is now
 \eq{.5.8.}{
\qrho \dot{\qini\qe} = - \qcurr{\qj}{\qini\qe}' + \qini\qsig \dot \qeps
 }
where \m { \qcurr{\qj}{\qini\qe} } is the heat current and prime denotes
the spatial derivative.

For the entropy current \m { \qcurr{\qj}{\qini\qs} }, we take the
standard choice \mm {\qcurr{\qj}{\qini\qs} = {\qj_{\qini\qe}}/{\qT} }
\cite{GroMaz62b}. Then, utilizing \re{.5.5.} in the balance for
entropy,
 \eq{.5.9.}{
\qrho \dot{\qini\qs} = - \qcurr{\qj}{\qini\qs}' + \qentprod{\qini\qs} ,
 }
gives for the entropy production \m { \qentprod{\qini\qs} }, via
\re{.5.6.} and \re{.5.8.},
 \eq{.5.10.}{
\qentprod{\qini\qs} = \qrho \dot{\qini\qs} + \qcurr{\qj}{\qini\qs}' =
\f{\qrho}{\qT} \qini{\dot\qe} - \f{\qini\qsig}{\qT} \dot \qeps + \9 1 {
\f{\qcurr{\qj}{\qini\qe}}{\qT} }' = -\f {\qcurr{\qj}{\qini\qe}'}{\qT} +
\9 1 { \f{\qcurr{\qj}{\qini\qe}}{\qT} }' = \qcurr{\qj}{\qini\qe}
 \0 1 { \f {1}{\qT} }' ,
 }
which we can ensure to be positive definite by choosing Fourier heat
conduction,
 \eq{.5.11.}{
\qcurr{\qj}{\qini\qe} = \qlam \0 1 { \f {1}{\qT} }' ,
 }
with a positive heat conduction coefficient \m { \qlam }.

We can also work in the canonical thermodynamical variables \m
{\qini\qe, \qeps }. This is achieved by expressing \m { \qT } from
\re{.4.4.} as a function of \m { \qini\qe } and \m { \qeps }, and
substituting it into the entropy, obtaining \m { \qs \1 1 {\qini\qe,
\qeps} }. For example, in the constant specific heat case, we have
 \eq{.5.12.}{
\qini\qs \1 1 {\qini\qe, \qeps} = \qc \ln \f { \qini\qe -
\F001{\qyoung}{2}\qeps^2}{\qc \qT_0} + \qini\qs_0 .
 }
In the canonical variables, temperature and stress are accessed \1 2
{cf.~\re{.5.6.}} as
 \eq{.6.13.}{
\f {1}{\qT} \1 1 { \qini\qe, \qeps } = \restr{ \f {\pd \qini\qs}{\pd
\qini\qe} }{\qeps} ,
 \qquad
\f {\qini\qsig}{\qrho \qT} \1 1 { \qini\qe, \qeps } = - \restr{ \f {\pd
\qini\qs}{\pd \qeps} }{\qini\qe} .
 }
Entropy can be shown to be concave in the canonical variables under the
conditions \mm { \qT > 0 ,}  \mm { \qyoung > 0 } and \mm { \F000{\dd
\qth\qe}{\dd \qT} > 0 .}

We note that the reason we started using the variables \m { \qT, \qeps }
instead of the canonical variables \m { \qini\qe, \qeps } is that this
enabled us to express the neglection of thermal expansion easily \1 2
{cf.~\re{.4.4.}}.

The first stage, the characterization of the initial system has thus
been completed.

\ssect{.6..2.2.}{Introducing the internal variable}

Now comes the second step: in addition to the variables \m {\qini\qe,
\qeps }, let us assume the existence of an additional variable \m { \qxi
}, and shift entropy by a \m { \qxi } dependent term. This term should
be concave and vanishing for \m { \qxi= 0 } \1 1 {in equilibrium}.
Assuming that the second derivative of entropy with respect to \m { \qxi
} is nonzero, by the Morse lemma, we can simply take the additional term
in the form \m { - \f {1}{2} \qxi^2 }:
 \eq{.6.14.}{
\qext{\qs} \1 1 {\qini\qe, \qeps, \qxi} = \qini\qs \1 1 {\qini\qe,
\qeps} - \f {1}{2} \qxi^2 ;
 }
namely, we can \textit{choose} \m { \qxi } to be the variable in which
the additional term is of this form. Note that this choice can be made
only if we have no direct physical knowledge about the origin of the
assumed internal variable and have thus a freedom in choosing it. If we
had some explicit information about \m { \qxi }---for example, a
microscopic interpretation---then we should not have enforced this
specific form for the additional term but a more general concave
function must have been allowed, possibly dependent on \m { \qini\qe  }
and \m{ \qeps } as well \cite{Ver97b,CiaVer93a}. In our present case,
we do not possess such background knowledge.

In the light of \re{.5.6.}, we find the Gibbs relation for the extended
entropy to be
 \eq{.6.15.}{
\qrho \dd \qext\qs = \f{\qrho}{\qT} \dd \qini\qe -
\f{\qini\qsig}{\qT} \dd \qeps - \qrho \qxi \dd \qxi .
 }
Most significantly, rheological effects manifest themselves in the
mechanical behavior so, in parallel, we allow an additional source of
mechanical stress also:
 \eq{.6.16.}{
\qext{\qsig} = \qini{\qsig} + \qirr{\qsig} ,
 }
\m {\qirr\qsig} being the stress contribution of rheological \1 1
{nonequilibrium} origin, and \m { \qext{\qsig} } denoting the total
stress. Consequently, in the mechanical power,
and in the balance for internal energy, an additional term \m {
{\qirr\qsig} \dot \qeps } appears:
 \eq{.6.17.}{
\qrho \dot{\qini\qe} + \qcurr{\qj}{\qini\qe}' = \qext{\qsig} \dot \qeps
= \qini{\qsig} \dot \qeps + {\qirr\qsig} \dot \qeps .
 }
Utilizing this and \re{.6.15.}, the entropy production can be calculated,
{assuming \m {\, \qcurr{\qj}{\qext\qs} = {\qj_{\qini\qe}}/{\qT} }}:
 \eq{.7.18.}{
\qentprod{\qext\qs} = \qrho \dot{\qext\qs} + \qcurr{\qj}{\qext\qs}' =
\f{\qrho}{\qT} \dot{\qini\qe} - \f{\qini\qsig}{\qT} \dot\qeps - \qrho
\qxi \dot \qxi + \9 1 { \f{\qcurr{\qj}{\qini\qe}}{\qT} }' = -\f
{\qcurr{\qj}{\qini\qe}'}{\qT} + \f{\qirr\qsig}{\qT} \dot \qeps - \qrho
\qxi \dot\qxi + \9 1 { \f{\qcurr{\qj}{\qini\qe}}{\qT} }' =
\qcurr{\qj}{\qini\qe} \cdot \0 1 { \f {1}{\qT} }' + \f{\qirr\qsig}{\qT}
\dot \qeps - \qrho \qxi \dot\qxi .
 }
In the rhs, for the first term (heat conduction), let us keep the
previous Fourier choice \m { \qcurr{\qj}{\qe} = \qlam \0 1 { \f {1}{\qT}
}' }; actually, in the three dimensional treatment we will see that in
isotropic materials heat conduction cannot couple to the rheological
side because the former is vectorial while the latter is a sum of a
tensorial (deviatoric) and a scalar (spherical) contribution.
Concerning the remaining part, let us ensure its positive
definiteness, rewritten as
 \eq{.7.19.}{
\qirr\qsig \dot\qeps - \qrho \qT \qxi \dot\qxi \ge 0 ,
 }
via Onsagerian equations
 \eq{.7.20.}{
\qirr\qsig \s= \ql_{11} \dot \qeps + \ql_{12} \0 1 {-\qrho \qT \qxi},
 \lel{.7.21.}
\dot \qxi \s= \ql_{21} \dot \qeps + \ql_{22} \0 1 {-\qrho \qT \qxi}
 }
with appropriate conditions on the coefficients \m { \ql_{ij} }.
Remarkably, these coefficients may depend on the state variables and
also on the thermodynamic forces. Therefore, the above consitutive
equations may be quasilinear and even nonlinear according to the
classification in \cite{Gya77a}.
 The conditions on the \m { \ql_{ij} }s are defined by the positive
definiteness of the quadratic form obtained by substituting
\re{.7.20.}--\re{.7.21.} into \re{.7.19.}, finding
 \eq{.7.22.}{
\ql_{11} {\dot \qeps}^2 + \0 1 { \ql_{12} + \ql_{21} }
\dot\qeps \0 1 {-\qrho \qT \qxi} + \ql_{22} \0 1 {-\qrho \qT \qxi}^2
 \s=
\0 1 { \mat{  \dot \qeps  &  -\qrho \qT \qxi  } }
\0 1 { \mat{  \ql_{11}  &  \ql_{12}\symm \\ \ql_{12}\symm  &  \ql_{22}  } }
\0 1 { \mat{  \dot \qeps  \\  -\qrho \qT \qxi  } }
\ge 0
 }
with \m { \ql_{12}\symm = \f {1}{2} \0 1 { \ql_{12} + \ql_{21} } }, the
offdiagonal element of the symmetric part of the matrix \m { \ql }.
Hence, the symmetric part \mm { \ql\symm } of the matrix \m { \ql } is
required to be positive definite. This necessitates, due to Sylvester's
criterion,
 \eq{.7.23.}{
\ql_{11} \ge 0,  \qquad  \ql_{22} \ge 0,  \qquad  \det \ql\symm \ge 0 .
 }
\1 1 {These three conditions are not independent: in fact, either of the
first two is implied by the other two.}

Note that the antisymmetric part, \m {\ql\asymm }, of the coefficient
matrix \m { \ql } does not contribute to the entropy production, it is
only the symmetric part that creates irreversibility.

\ssect{.7..2.3.}{Eliminating the internal variable}

As mentioned above, the coefficients \m { \ql_{ij} } need not be
constants, and their dependence in the present case could be on
temperature. Now let us assume that \m { \ql_{11} }, \m { \qrho \qT
\ql_{12} }, \m { \ql_{21} } and \m { \qrho \qT \ql_{22} } are constant,
at least along a process, at least to a good approximation---{which is a
very frequent situation}.  In this case it is easy to eliminate the
internal variable \m { \qxi } from \re{.7.20.}--\re{.7.21.} (although
the elimination is straightforward at the general level, too). Notably,
the elimination of internal variables has already been applied by
Meixner, in \cite{Mei54a}.

We start by rewriting \re{.7.21.} in a form where a differential
operator acts on \m { \qxi }:
 \eq{.8.24.}{
\0 1 { \qD + \qrho \qT \ql_{22} } \qxi = \ql_{21} \dot \qeps .
 }
Then, if we operate \m { \qD + \qrho \qT \ql_{22} } on \re{.7.20.},
utilizing \re{.8.24.} yields
 \eq{.8.25.}{
\qrho \qT \ql_{22} \qirr\qsig + \dot{\qirr\qsig} \s= \qrho \qT (\det \ql)
\bitt \dot \qeps + \ql_{11} \ddot \qeps .
 }
Knowing \m { \qirr\qsig = \qext\qsig - \qyoung \qeps }, we can bring
this into the final form
 \eq{.8.26.}{
\qext\qsig + \f {1}{\qrho \qT \ql_{22}} \dot{\qext\qsig} \s = \qyoung
\qeps +  \0 1 { \f {\det \ql}{\ql_{22}} + \f {\qyoung}{\qrho \qT
\ql_{22}} } \dot \qeps + \f {\ql_{11}}{\qrho \qT \ql_{22}} \ddot \qeps .
 }
That is, we have obtained a \model{0, 1}{0, 1, 2} rheological model:
 \eq{.8.27.}{
\qext\qsig + \qtau \dot{\qext\qsig} \s = \qalp \qeps + \qbet \dot \qeps
+ \qgam \ddot \qeps ,
 }
where
 \eq{.8.28.}{
\qtau \s = \f {1}{\qrho \qT \ql_{22}} > 0 ,
 &
\qalp \s = \qyoung ,
 &
\qbet \s = \f {\det \ql}{\ql_{22}} + \f {\qalp}{\qrho \qT \ql_{22}}
\ge \f {\qalp}{\qrho \qT \ql_{22}} > 0 ,
 &
\qgam \s = \f {\ql_{11}}{\qrho \qT \ql_{22}} \ge 0 ,
 }
which are the necessary and sufficient conditions coming from
thermodynamics.

A minor inconvenience is that \m { \qtau = 0 } and \m { \qbet = 0 } are
excluded so the \model{0}{0} model, \m { \qext\qsig = \qalp \qeps } is
not included dirctly but can be covered as a \m { \ql_{22} \to \infty }
limit. We can easily overcome this nuisance. Let us observe that,
assuming \m { \ql_{22} > 0 }, it is possible to rearrange
\re{.7.20.}--\re{.7.21.} as
 \eqa{.8.29.}{
\qirr\qsig \s = \f {\det \ql}{\ql_{22}} \dot \qeps + \f
{\ql_{12}}{\ql_{22}} \dot \qxi = \qm_{11} \dot \qeps + \qm_{12} \dot \qxi ,
 \lel{.8.30.}
\0 0{- \qrho \qT \qxi} \s = - \f {\ql_{21}}{\ql_{22}} \dot \qeps + \f
{1}{\ql_{22}} \dot\qxi = \qm_{21} \dot \qeps + \qm_{22} \dot \qxi .
 }
Now, equations of the form
 \eqa{.8.31.}{
\qirr\qsig \s = \qm_{11} \dot \qeps + \qm_{12} \dot \qxi ,
 \lel{.8.32.}
\0 0{- \qrho \qT \qxi} \s = \qm_{21} \dot \qeps + \qm_{22} \dot \qxi
 }
are just another possible valid Onsagerian way to ensure the positive
definiteness of \re{.7.19.}. Namely, using \re{.8.31.}--\re{.8.32.},
\re{.7.19.} has the form
 \eq{.8.33.}{
\qm_{11} {\dot \qeps}^2 + \0 1 { \qm_{12} + \qm_{21} }
\dot\qeps \dot \qxi + \qm_{22} {\dot \qxi}^2 =
\0 1 { \mat{  \dot \qeps  &  \dot \qxi  } }
\0 1 { \mat{  \qm_{11}  &  \qm_{12}\symm
 \\
\qm_{12}\symm  &  \qm_{22}  } }
\0 1 { \mat{  \dot \qeps  \\  \dot \qxi  } } ,
 }
again a quadratic form which is positive definite if and only if
 \eq{.9.34.}{
\qm_{11} \ge 0,  \qquad  \qm_{22} \ge 0,  \qquad  \det \qm\symm \ge 0 .
 }
Note again that only the symmetric part of \m { \qm } is related to the
entropy production.

When we eliminate \m { \qxi } from \re{.8.31.}--\re{.8.32.}---now
observing the differential operator \mm { \qm_{22} \qD + \qrho \qT } in
the time derivative of \re{.8.32.}, and applying this operator on
\re{.8.31.}---, the result is
 \eq{.9.35.}{
\qext\qsig + \f {\qm_{22}}{\qrho \qT} \dot{\qext\qsig} \s= \qyoung \qeps +
\0 1 { \qm_{11} + \f {\qm_{22}}{\qrho \qT} \qyoung }
\dot \qeps + \f {\det \qm}{\qrho \qT} \ddot \qeps .
 }
The coefficients are thus
 \eq{.9.36.}{
\qtau \s = \f {\qm_{22}}{\qrho \qT} \ge 0 ,
 &
\qalp \s = \qyoung ,
 &
\qbet \s = \qm_{11} + \f {\qm_{22}}{\qrho \qT} \qalp
\ge 0 ,
 &
\qgam \s =\f {\det \qm}{\qrho \qT} \ge 0 .
 }
We again obtain essentially the same family of rheological models,
differences being only at the boundary of the thermodynamically allowed
parameter region. Indeed, the latter way, the \model{0}{0} Hooke model
is also incorporated. Moreover, the \m { \qtau = 0 }, \m { \qalp = 0 },
\m { \qbet = 0 }, \m { \qgam > 0 } model---in other words, the
\model{0}{2} body---\1 1 {on which see more in Sect.~\ref{.9..2.4.}} is
also uncovered as a thermodynamically valid case. In parallel, there is
a part of the boundary that is missing here but was allowed in the
former formulation: \m { \ql_{22} = 0 } would have allowed \model{1}{1,
2} models, too. However, those relationships between \m { \qsig }, \m {
\qeps } and derivatives do not include the \model{0}{0} Hooke case, do
not provide information for static processes of solids and are thus
incomplete {for solids}.

Hereafter, for definiteness, the coefficients \m { \qm_{ij} } will be
used but analogous statements will hold for the coefficients \m
{\ql_{ij} } as well.

\ssect{.9..2.4.}{Special cases and analysis}

The obtained \model{0, 1}{0, 1, 2} rheological model \re{.8.27.} covers a
number of well-known classic cases. In fact, it includes the
\model{0}{0} Hooke model, the \model{0}{0, 1} Kelvin--Voigt model, the
{\model{0, 1}{1}} Maxwell model, the {\model{0, 1}{0, 1}}
Poynting--Thomson--Zener model, and the {\model{0, 1}{1, 2}} Jeffreys
model. In this respect, an important contribution of the present
internal state variable approach is that it reveals thermodynamical
constraints of the coefficients of these models. Based on \re{.9.34.},
\re{.9.36.} and the identity \eq{.9.37.}{ \det \qm = \det \qm\symm + \0 1
{ \f {\qm_{12} - \qm_{21}}{2} }^2 = \det \qm\symm + \0 1 {\qm\asymm_{12}
}^2, } these constraints read
 \eq{.9.38.}{
 \qquad
 \qquad
\qtau \s \ge 0 ,
 &
\qalp \s \ge 0 ,
 &
\qindb \s := \qbet - \qtau \qalp = \qm_{11} \ge 0 ,
 &
\qgam \s \ge 0 ,
 \qquad
 \qquad
 }
where the \textit{index of damping}, \m { \qindb }, has been introduced.
This combination \m { \qindb }, together with the similarly defined
\textit{index of inertia},
 \eq{.10.39.}{
\qindc := \qgam - \qtau \qindb = - \f {\qm_{12}
\qm_{21}}{\qrho \qT} = \f { \0 1 { \qm_{12}\asymm }^2 - \0 1 {
\qm_{12}\symm }^2 }{\qrho \qT} ,
 }
are two important characteristics of the {\model{0, 1}{0, 1, 2}} models.
\m { \qindb \ge 0 } means a nontrivial combined condition for \m {\qtau
}, \m { \qalp } and \m { \qbet }, which, for example, rules out the
existence of \model{0, 1}{0} and \model{0, 1}{0, 2} models, on
thermodynamical ground. In parallel, \m { \qindc } is allowed to be
positive as well as negative but these two cases indicate remarkably
different situations. When \m { \qindc < 0 }, in other words, when the
symmetric part of the coefficient matrix \m { \qm } dominates over the
antisymmetric part, then if stress is prescribed as a function of time
then the characteristic equation of the emerging second order linear
inhomogeneous differential equation \1 2 {\m { \qeps_{\txt00{homog.}} \1
1 {t} \sim \e^{\qlam t} \; \Rightarrow \; \qalp + \qbet \qlam + \qgam
\qlam^2 = 0 }} has two negative real roots, meaning that the solution is
characterized by two decreasing exponential functions of time. On the
other side, when \m { \qindc > 0 }, \ie when the antisymmetric part
dominates, the two roots are not real. The real parts are negative,
ensuring damping, but the imaginary parts describe oscillations in the
solution. This opens the possibility for bringing the material into
resonance via a periodic excitation with a frequency near to the
self-frequency of the rheological material. It is important to note that
this resonance is completely different from elastic resonance, which
occurs when the geometric sizes of a body, the velocity of elastic waves
within the body, and the frequency of the periodic external force at the
boundary, are in tune. The rheological resonance is a completely local
phenomenon, independent from the geometrical properties of the body, and
is a result of the relationship among \m {\qalp} and the two rheological
coefficients \m { \qbet } and \m { \qgam }. The inertia-like coefficient
refers not to the usual mechanical inertia but to a rheology-related
different type of inertia.

That the \textit{overdamped} (or \textit{underinertial}), \ie \m {\qindc
< 0}, models differ remarkably from the
\textit{underdamped/overinertial} (\m {\qindc > 0}) cases can be
demonstrated in another way as well. Namely, one can show
\cite{Ful08c}---via steps
analogous to those in Appendix A---that, when two
rheological networks are arranged in parallel, the sum of their index of
damping equals the index of damping of the resulting system, and the
same additive property holds for the index of inertia as long as the
resulting model is not beyond \model{0, 1}{0, 1, 2}.
Furthermore, when two rheological networks are arranged in series,
non-negative indices of damping also prove to lead to non-negative index
of damping (and positive ones to positive), and nonpositive indices of
inertia lead to nonpositive index of inertia (and negative ones to
negative). A reassuring consequence is that networks created by springs
and dashpots in a combination of serial and parallel arrangements always
have positive index of damping, in conform with the thermodynamical
constraint. A nontrivial other consequence, however, is that springs and
dashpots---possessing zero index of inertia---can never be combined, in
any serial-plus-parallel way, to a rheological model with positive index
of inertia. In particular, the generalized Kelvin--Voigt and generalized
Maxwell--Wiechert models---which are Kelvin--Voigt or Maxwell models
in serial, resp.\ parallel, arrangement---are not able to reproduce
these thermodynamically completely legitim models.

This necessitates the introduction of a new rheological element,
corresponding to the \model{0}{2} model. The need for this new element
was indicated first---to our knowledge--- by Verh\'as \cite{Ver85b,Ver97b}.
Hereafter, we call this element the \emp{inertial} element,
and its depicting will follow Verh\'as' notation (see Fig.~1).

 \begin{figure}
\newcounter{caa}  \setcounter{caa}{37}  
\newcounter{cab}  \setcounter{cab}{\thecaa}  \addtocounter{cab}{\thecaa}
\newcounter{cac}  \setcounter{cac}{10}  
\newcounter{cad}  \setcounter{cad}{35}
\newcounter{cae}  \setcounter{cae}{15}
\newcounter{caf}  \setcounter{caf}{\thecaa}  \addtocounter{caf}{\thecae}
\newcounter{cag}  \setcounter{cag}{\thecaa}  \addtocounter{cag}{-\thecae}
\newcounter{cah}  \setcounter{cah}{3}
\newcounter{cai}  \setcounter{cai}{\thecac}  \addtocounter{cai}{\thecah}
\newcounter{caj}  \setcounter{caj}{\thecac}  \addtocounter{caj}{-\thecah}
\newcounter{cak}  \setcounter{cak}{50}  
  \begin{center}\begin{picture}(\thecab,\thecad)(0,0)
\linethickness{.12ex}  
\put(\thecaf,\thecaj){\vector(-1,0){\thecak}}  
\put(\thecag,\thecai){\vector(1,0){\thecak}}
\put(\thecaa,29){\makebox(0,0){$\qgam$}}
\thicklines  
\put(\thecaa,\thecac){\circle{6}}  \put(\thecaa,\thecac){\circle{20}}
 \end{picture}\end{center}
\caption{The inertial element, corresponding to the elementary
\model{0}{2} model, \m { \qext\qsig = \qgam \ddot \qeps }}
 \end{figure}

\sect{.11..3.}{The procedure in three spatial dimensions}

Now we treat the three dimensional case, in complete analogy to the
one dimensional situation.

\ssect{.11..3.1.}{The initial system}

In three dimensions, elastic strain, \m { \qqeps }, and stress, \m{
\qini\qqsig }, are symmetric tensors. Assuming isotropic Hooke
elasticity, stress depends linearly on elastic strain via two scalar
elastic coefficients, one connecting the deviatoric tensorial components
and the other relating the spherical ones:
 \eqa{.11.40.}{
\qini\qqsig \1 1 {\qqeps} \s=
\qyoung\dev \qqeps\dev + \qyoung\sph \qqeps\sph ,
\quad \quad  \qqeps\sph =\f {1}{3}\1 1 { \tr \qqeps } \tens1 ,
\quad  \qqeps\dev = \qqeps - \qqeps\sph ,
\quad  \qyoung\dev = 2G ,
\quad  \qyoung\sph = 3K .
 }
We stay in the small-deformation regime, where \1 1{approximately} the
density \m { \qrho } is constant and the velocity gradient tensor
equals \m { \dot \qqeps }. Correspondingly, the mechanical power of \m {
\qini\qqsig } can be expressed as
 \eq{.11.41.}{
\tr \1 1 { \qini\qqsig \dot \qqeps } = \tr \1 1 { \qini{\qqsig\dev}
{\dot\qqeps}\dev } + \tr \1 1 { \qini{\qqsig\sph} {\dot\qqeps}\sph } =
\qrho \qel{\dot{\qe}}
 }
with
 \eq{.11.42.}{
\qel{\qe} \1 1 { \qqeps } = \f {\qyoung\dev}{2\qrho} \tr \0 1 { \0 0
{\qqeps\dev} {\qqeps\dev} } + \f {\qyoung\sph}{2\qrho} \tr \0 1 { \0 0
{\qqeps\sph} {\qqeps\sph} } .
 }
Total {specific} energy is considered in the form
 \eq{.11.43.}{
\qini\qe \1 1 {\qT, \qqeps} = \qth{\qe} \1 1 {\qT} + \qel\qe \1 1
{\qqeps} .
 }
The Gibbs relation \mm{ \qrho \dd \qini\qe = \qrho \qT \dd \qini\qs +
\tr \1 1 {\qini\qqsig \dd \qqeps} ,} rearrangable as
 \eq{.11.44.}{
\qrho \dd \qini\qs = \f{\qrho}{\qT} \dd \qini\qe -
\f{1}{\qT} \tr \1 1 {\qini\qqsig \dd \qqeps} ,
 }
holds with a {specific} entropy \m { \qini\qs = \qini\qs \1 1 {\qT} }
satisfying
 \eq{.12.45.}{
\f {\dd \qini\qs}{\dd \qT} = \f {1}{\qT} \f{\dd \qth\qe}{\dd \qT} .
 }
With the expression of power \re{.11.41.}, the balance of internal energy
\1 1 {the first law} is
 \eq{.12.46.}{
\qrho \dot{\qini\qe} = - \nabla \cdot \qcurr{\qqj}{\qini\qe} + \tr \1 1
{\qqsig \dot{\qqeps}} .
 }
We take the standard choice \mm {\qcurr{\qqj}{\qini\qs} =
{\qqj_{\qini\qe}}/{\qT} } between the entropy current \m {
\qcurr{\qqj}{\qini\qs} } and the heat current \mm
{\qcurr{\qqj}{\qini\qe} }.
 Then, in the balance for entropy,
 \eq{.12.47.}{
\qrho \dot{\qini\qs} = - \nabla \cdot \qcurr{\qqj}{\qini\qs} +
\qentprod{\qini\qs} ,
 }
the entropy production \m { \qentprod{\qini\qs} } is calculated, using
\re{.11.44.} and \re{.12.46.}, as
 \eq{.12.48.}{
\qentprod{\qini\qs} = \qrho \dot{\qini\qs} + \nabla \cdot
\qcurr{\qqj}{\qini\qs} = \f{\qrho}{\qT} \qini{\dot\qe} - \f{1}{\qT} \tr
\1 1 {\qqsig \dot{\qqeps}} + \nabla \cdot \9 1 {
\f{\qcurr{\qqj}{\qini\qe}}{\qT} } = - \f {1}{\qT} \nabla \cdot
\qcurr{\qqj}{\qini\qe} + \nabla \cdot \9 1 {
\f{\qcurr{\qqj}{\qini\qe}}{\qT} } = \qcurr{\qqj}{\qini\qe} \cdot \nabla
\0 1 { \f {1}{\qT} } ,
 }
which we set positive definite via choosing Fourier heat conduction,
 \eq{.12.49.}{
\qcurr{\qj}{\qini\qe} = \qlam \nabla \0 1 { \f {1}{\qT} } ,
 \qquad  \qlam > 0 .
 }

The conversion from state variables \m { \1 1 { \qT, \qqeps } } to the
canonical state variables \m { \1 1 { \qini\qe, \qqeps } } can be
achieved by expressing \m { \qT } from \re{.11.43.}, and substituting it
into the entropy, obtaining \m { \qs \1 1 {\qini\qe, \qqeps} }.
Reversely, starting from the canonical variables, temperature is
accessed as
 \eq{.12.50.}{
\f {1}{\qT} \1 1 { \qini\qe, \qqeps } = \restr{ \f {\pd \qini\qs}{\pd
\qini\qe} }{\qqeps} .
 }
Entropy is concave in the canonical variables \1 2 {under the natural
conditions \mm { \qT > 0 ,} \mm { \qyoung > 0 } and \m { \, \F000{\dd
\qth\qe}{\dd \qT} > 0 }}.

Having the initial system in the canonical variables, we are ready for
extending it.

\ssect{.12..3.2.}{Introducing the internal variable}

The extended thermodynamical state space is chosen to be spanned by the
following variables: internal energy \m { \qini\qe }, strain \m { \qqeps
}, and an internal variable \m{ \qqxi }. This latter is taken as a
second order symmetric tensor, based on our purpose to gain an extension
of the mechanical aspects \1 1 {the \quot{material law}} of the initial
system, to obtain corrections to the relation between stress and strain,
which quantities are both symmetric tensors.

We shift entropy by a concave nonequilibrium term
depending---quadratically---on \m { \qqxi } only. According to the Morse
lemma, this new entropy term can be chosen as a pure square term, hence,
the extended entropy function is
 \eq{.12.51.}{
\qext{\qs} \1 1 {\qini\qe, \qqeps, \qqxi} =
\qini\qs \1 1 {\qini\qe, \qqeps} - \f {1}{2} \tr \0 1 { \qqxi^2 } .
 }
The Gibbs relation for the extended entropy is
 \eq{.13.52.}{
\qrho \dd \qext\qs = \f{\qrho}{\qT} \dd \qini\qe - \f{1}{\qT} \tr \1 1
{\qini\qqsig \dd \qqeps} - \qrho \tr \0 1 { \qqxi \dd \qqxi } .
 }

To obtain some effect on the mechanical aspects, stress is also
considered extended by a rheological \1 1 {nonequilibrium} term:
 \eq{.13.53.}{
\qext{\qqsig} = \qini{\qqsig} + \qirr{\qqsig} .
 }
Consequently, the mechanical power \re{.11.41.}, and correspondingly the
energy balance, gets shifted as
 \eq{.13.54.}{
\qrho \dot{\qini\qe} + \nabla \cdot \qcurr{\qqj}{\qini\qe} = \tr \1 1 {
\qext\qqsig \dot \qqeps } = \tr \1 1 { \qini\qqsig \dot \qqeps } + \tr
\1 1 { \qirr\qqsig \dot \qqeps } .
 }
With the choice \mm {\qcurr{\qqj}{\qext\qs} = {\qqj_{\qini\qe}}/{\qT} ,}
and utilizing \re{.13.52.} and \re{.13.54.}, the entropy production is
found to be
 \eqn{@145.}{
\qentprod{\qext\qs} \s= \qrho \dot{\qext\qs} + \nabla \cdot
\qcurr{\qqj}{\qext\qs} = \f{\qrho}{\qT} \dot{\qini\qe} - \f{1}{\qT} \tr
\1 1 {\qqsig \dot{\qqeps}} - \qrho \tr \2 1 { \qqxi \dot \qqxi } +
\nabla \cdot \9 1 { \f{\qcurr{\qqj}{\qini\qe}}{\qT} }
 \lel{.13.55.}
\s= - \f {1}{\qT} \nabla \cdot \qcurr{\qqj}{\qini\qe} + \f {1}{\qT} \tr
\1 1 { \qirr\qqsig \dot \qqeps } - \qrho \tr \2 1 { \qqxi \dot \qqxi }
+ \nabla \cdot \9 1 { \f{\qcurr{\qqj}{\qini\qe}}{\qT} }
 \lel{.13.56.}
\s= \qcurr{\qqj}{\qini\qe} \cdot \nabla  \0 1 { \f {1}{\qT} } +
 \f{1}{\qT} \tr \2 1 { \qirr\qqsig\dev \dot \qqeps\dev }
+ \f{1}{\qT} \tr \2 1 { \qirr\qqsig\sph \dot \qqeps\sph }
- \qrho \tr \2 1 { \qqxi\dev \dot \qqxi{}\dev }
- \qrho \tr \2 1 { \qqxi\sph \dot \qqxi{}\sph } .
 }
In the rhs, vectors are present in the first term, scalars in the third
and fifth one, and symmetric traceless tensors in the second and fourth
term. In an isotropic material, these three types of quantities cannot
couple to one another according to the Curie principle, the
representation theorem of isotropic functions \cite{Ver97b}. Therefore,
concerning the term containing
vectors, we consider Fourier heat conduction, \mm { \qcurr{\qqj}{\qe} =
\qlam \nabla \0 1 { \f {1}{\qT} } }. For the remaining two pairs of
terms, the most general Onsagerian solution is
 \eq{.13.57.}{
 \qquad
\qirr\qqsig\dev \s= \ql_{11}\dev \dot \qqeps\dev +
\ql_{12}\dev \2 1 {\mathord- \qrho \qT \qqxi\dev} ,
 &
\qirr\qqsig\sph \s= \ql_{11}\sph \dot \qqeps\sph +
\ql_{12}\sph \2 1 {\mathord- \qrho \qT \qqxi\sph} ,
 \qquad
 \lel{.13.58.}
 \qquad
\dot \qqxi{}\dev \s= \ql_{21}\dev \dot \qqeps\dev +
\ql_{22}\dev \2 1 {\mathord- \qrho \qT \qqxi\dev} ,
 &
\dot \qqxi{}\sph \s= \ql_{21}\sph \dot \qqeps\sph +
\ql_{22}\sph \2 1 {\mathord- \qrho \qT \qqxi\sph} ,
 }
or the corresponding version with \m { \qm\dev } and \m { \qm\sph },
generalizing \re{.8.29.}--\re{.8.30.}.

Hence, we can see that what we had in one dimension just gets doubled in
three dimensions, the two components being independent. The two terms
with traceless symmetric tensors have to be positive definite
themselves, and the two terms with scalars have to be positive definite
independently. The corresponding conditions on the coefficients are the
same as for the one dimensional case.

Eliminating the internal variable in the constant temperature case also
leads to two independent \model{0, 1}{0, 1, 2} models,
 \eq{.13.59.}{
\qqsig\dev + \qtau\dev \dot\qqsig\dev \s = \qalp\dev \qqeps\dev +
\qbet\dev \dot \qqeps\dev + \qgam\dev \ddot \qqeps\dev ,
 &
\qqsig\sph + \qtau\sph \dot\qqsig\sph \s = \qalp\sph \qqeps\sph +
\qbet\sph \dot \qqeps\sph + \qgam\sph \ddot \qqeps\sph ,
 }
with thermodynamics-originated inequalities for the altogether eight
coefficients.

Remarkably, in a uniaxial process, both the deviatoric and the spherical
rheologies are active, independently of each other, and cause a rather
complicated resultant uniaxial rheology---see Appendix~\ref{.16..A.} for
the details.

We close this section with two notes. The first is that the presented
thermodynamical approach involves a potential (entropy, and free energy as
a consequence), therefore, our treatment is hypereleastic---from the
point of view of the complete state space. On the other side, the
stress--strain relation obtained by reduction/elimination cannot be
obtained from a potential.

The second remark is that more general methods of the exploitation
of the entropy principle \cite{Van03a} result, in this classical
irreversible thermodynamical case, in the same structure as the simple
approach of divergence separation applied here.

\sect{.14..4.}{Comparison of the approaches}

Kluitenberg \cite{Klu63a} has also obtained a result similar to ours, a
pair of \model{0, 1}{0, 1, 2} models, with assumptions different from
ours, and one can also derive the \model{0, 1}{0, 1}
Poynting--Thomson--Zener body in Extended Irreversible Thermodynamics
\cite{DauLeb90a}. These approaches and their different assumptions are
discussed in the following section.

\ssect{.14..4.1.}{Kluitenberg theory}

We have seen that the presented thermodynamical framework distinguishes
a particular combination of inertia, relaxation and creep for
dissipation, the \model{0, 1}{0, 1, 2} model, as a basic rheological
body. Nonequilibrium thermodynamics with internal variables was first
systematically applied to linear viscoelasticity by Kluitenberg
\cite{Klu62a1,Klu62a2,Klu62a3,Klu63a,Klu64a,Klu68a,KluCia78a}, and he
obtained also the \model{0, 1}{0, 1, 2} body,
as a fundamental building block of thermodynamical modelling.

The Kluitenberg theory is different from our approach. The differences
are the following:
  \begin{enumerate}
 \item
In the Kluitenberg theory, the internal variable has a fixed physical
meaning: it is interpreted as a direct---called anelastic---contribution
to the strain: \cite{Klu62a1}
 \eq{.14.60.}{
\qqeps = \qel\qqeps + \qanel\qqeps.
 }
According to our above approach, this interpretation is not necessary,
the coupling of the dissipative terms of the entropy production through
linear relations with an arbitrary tensorial internal variable
influences the strain.
 \item
Kluitenberg assumes a relationship between anelastic and elastic
stress, as a consequence of the strain interpretation. In particular,
for him, the elastic stress is equal to the anelastic one
\1 3 {\cite{Klu62a3}, (4.4)}. He assumes, in particular, that
 \eq{.14.61.}{
\f{\partial \qirr\qs}{\partial \qanel\qqeps}(\qel\qqeps,\qanel\qqeps) =
\f{\partial \qs}{\partial \qanel\qqeps}(\qqeps,\qanel\qqeps)
-\f{\partial \qs}{\partial \qqeps}(\qqeps,\qanel\qqeps) = 0.
 }
Here, \mm { \qirr \qs(\qel\qqeps,\qanel\qqeps) = \qirr
\qs(\qqeps-\qanel\qqeps,\qanel\qqeps)= \qs(\qqeps,\qanel\qqeps) }.
 \end{enumerate}

As we have found above, these assumptions are not necessary for
thermodynamical consistency, and lead to a different analysis of the
rheological coefficients. For example, in the Kluitenberg
representation, no restriction is revealed on the sign of \m {\qbet},
the coefficient of the strain rate. Moreover, when elastic and anelastic
strains of Kluitenberg are quadratic in the entropy, one cannot derive
the \model{0, 1}{0, 1, 2} body but obtains the \model{0, 1}{1, 2}
Jeffreys model, instead \1 3 {\cite{Klu62a3}, (4.25)--(4.26)}. However,
the Jeffreys body is not a rheological solid (Jeffreys model expresses a
viscous fluid for slow processes), hence, it is unsatisfactory in case
of viscoelastic solids. The real basic building block is the \model{0,
1}{0, 1, 2} body.

Verh\'as introduced the procedure analogous to ours here, with a single
internal variable, for fluids \cite{Ver85b,Ver97b}. In addition, it was
him who emphasized the relevance of the inertial element on
thermodynamical grounds. To reflect these precursors, hereafter we will
call the \model{0, 1}{0, 1, 2} rheological model the \emp{\qkv\ body}.

\ssect{.15..4.2.}{Extended Thermodynamics}

In Extended Irreversible Thermodynamics, the dissipative fluxes are
considered as state variables \cite{JouAta92b}. Therefore, the viscous
or anelastic strain \m { \qirr\qqsig = \qext\qqsig -\qqsig} is a
thermodynamical state variable, instead of being a constitutive quantity
(see, \eg \cite{DauLeb90a}). In our approach, if \m {\ql_{11}=0}, then
the internal variable \m { \qqxi } is proportional to the dissipative
stress \m { \qirr\qqsig } \1 2 {see \re{.8.28.}}, thus a rescaling of \m
{ \qqxi } results in the representation and rheological equations of
Extended Thermodynamics. The rheological body obtained this way is the
\model{0, 1}{0, 1} Poynting--Thomson--Zener body, a subfamily of the
full \model{0, 1}{0, 1, 2} range of possibilities. It is remarkable that
the material parameter set to zero here, \m { \ql_{11} }, is the
viscosity of the Kelvin body in the sense of  \re{.7.20.}, which is
normally the main contribution to dissipation. In this sense, the
situation is similar to heat conduction, where the representation of the
Fourier coefficient is different in case of internal variables and in an
Extended Thermodynamical treatment \cite{VanFul12a}. The extension of the
theory with higher order fluxes is different from the extension of an
internal variable theory with further internal variables.

\sect{.15..5.}{Discussion}

In this paper, we have analysed rheological properties of solids as a
thermodynamical theory with internal variables.
We have obtained a simple non-linear viscoelastic theory.
We have identified the basic
thermorheological body and investigated its properties. The single
internal thermodynamical state variable representation distinguishes the
standard Poynting--Thomson--Zener body supplemented by an inertial
element as the fundamental building block of linear viscoelasticity. We
have suggested to name this material model the \qkv\ body.

On one hand, the simplest assumptions were introduced, when the internal
variable is a single tensorial dynamical degree of freedom, representing
the deviation from the equilibrium state and is subject to thermodynamical
restrictions. On the other hand, we have allowed the most general linear
Onsagerian equations. The advantage of this is the obtainable
\emp{universality}, in the sense that, as long as the tensorial
representation and the second law are respected, any particular
microscopic and mesoscopic mechanism must lead to the same rheological
model family in the linear regime.

After eliminating the internal variable, for isothermal processes, we
could formulate and analyse the consequences of the presence of rheology
in measurable quantities, stress and strain.

In our approach, the internal variable is not necessarily anelastic
strain like in the classical Kluitenberg theory, and is not necessarily
the dissipative stress like in Extended Thermodynamics. The latter
choice is found to be a special case, a certain subfamily within our
obtained family of models. The representation of the internal variable
as additive deviation from elastic strain would also require a deeper
analysis from the point of view of continuum kinematic quantities,
including finite strain as well as the requirements of material frame
indifference. In this respect, it is remarkable that our recent frame
free and objective approach \cite{FulVan12a}, which improves the
classic framework of Noll \cite{Nol67a,Nol06a}, leads to an additive
decomposition of the deformation rate into an elastic and a plastic
(permanent) part. Therefore, the basic question is the clear distinction
of the elastic, rheological (recoverable) and the plastic (permanent)
parts of the deformation in experiments.

Another characteristic property of our treatment was that we have not
assumed symmetry or antisymmetry of the phenomenological coefficient
matrix. There is no reason to assume Onsager--Casimir reciprocity
without any microscopic interpretation. Moreover, the detailed analysis
of section \re{.9..2.4.} has shown that the different possible circuit-like
spring--dashpot-based representations are related to different, either
symmetric or antisymmetric part dominated thermodynamical models.
Thermodynamics requires non-negative dashpot and spring coefficients of
both representations and the inertial element is also needed for the underdamped case.

Finally, we emphasize that all the found restrictive properties of our
model and the differences from the original Kluitenberg and Extended
Thermodynamical approaches can be tested by experiments. The comparison of
the coefficient restrictions of the Kluitenberg model with wave
propagation experiments has been started by Ciancio \etal\ in
\cite{CiaEta07a,Cia08a,CiaEta08a} with satisfactory results. Direct
measurements of creep and relaxation require a careful choice of
isotropic materials with a single internal variable and also the
separation of spherical and deviatoric parts of the changes. Mixed
loading conditions and effective models can be complicated, especially
if not only deviatoric but volumetric anelasticity also plays a role
(see Appendix A). In this respect, the borehole rock sample experiments
of Lin \etal\ \cite{MatTak93a,Mats08a,LinEta10p} support the relevance
of both deviatoric and volumetric \qkv\ bodies.

\sect{.16..6.}{Acknowledgement}

This work was supported by the Hungarian National Research Fund OTKA
under contracts K81161, K82024, K104260.

\appendix

\sect{.16..A.}{Uniaxial rheology derived from the deviatoric and
spherical components}

During uniaxial processes of solids, the stress and strain tensors are,
in a suitable Cartesian coordinate system, characterized by matrices
 \eqa{.17.62.}{
\qqsig \s= \0 1{ \smat{ \qsig\para & & \\ & 0 & \\ & & 0 } } ,
 &
\qqsig\sph \s= \f{1}{3} \0 1{ \smat{ \qsig\para & & \\
& \qsig\para & \\ & & \qsig\para } } ,
 &
\qqsig\dev \s= \f{1}{3} \0 1{ \smat{ 2 \qsig\para & & \\
& - \qsig\para & \\ & & - \qsig\para } } ,
 \lel{.17.63.} 
\qqeps \s= \0 1 { \smat{ \qeps\para & & \\ & \qeps\orth & \\
& & \qeps\orth } } ,
 &
\qqeps\sph \s= \f{1}{3} \0 1{ \smat{ \qeps\para + 2 \qeps\orth & & \\
& \qeps\para + 2 \qeps\orth & \\ & & \qeps\para + 2 \qeps\orth } } ,
 &
\qqeps\dev \s= \f{1}{3} \0 1{ \smat{ 2 \0 1{ \qeps\para - \qeps\orth }
& & \\ & - \0 1{ \qeps\para - \qeps\orth } & \\ & &
- \0 1{ \qeps\para - \qeps\orth } } } ,
 }
all omitted matrix elements being zero.

Let our solid obey the linear rheological laws
 \eqa{.17.64.}{
\Dsd \bitt \qqsig\dev \s = \Ded \bitt \qqeps\dev ,
 \lel{.17.65.}  
\Dss \bitt \qqsig\sph \s = \Des \bitt \qqeps\sph ,
 }
where the differential operators \m{\Dsd}, \m{\Dss}, \m{\Ded},
\m{\Des} are polynomials of the time derivative operator
\m{\F000{\dd}{\dd t}}, with constant coefficients. For example, for
\re{.13.59.},
 \eq{.17.66.}{
 &&
\Dsd \s= 1 + \qtau\dev \qD ,
 &
\Ded \s= \qalp\dev + \qbet\dev \qD + \qgam\dev \qDD ,
 &&
 \lel{.17.67.}
 &&
\Dss \s= 1 + \qtau\sph \qD ,
 &
\Des \s= \qalp\sph + \qbet\sph \qD + \qgam\sph \qDD .
 &&
 }
Then, for uniaxial processes, \re{.17.64.} simplifies to the scalar
equation
 \eq{.17.68.}{
\Dsd \bit \qsig\para \s= \Ded \bit \0 1{ \qeps\para - \qeps\orth } ,
 }
and \re{.17.65.} gives
 \eq{.17.69.}{
\Dss \bit \qsig\para \s= \Des \bit \0 1{ \qeps\para + 2 \qeps\orth } .
 }
Our aim is to eliminate \m { \qeps\orth } from this pair of equations.
Let the operator \m{2 \Des} act on \re{.17.68.}, let \m{\Ded} act on
\re{.17.69.}, and let us consider the sum of the two resulting equations.
Observing that
 \eq{.17.70.}{
\Des \Ded = \Ded \Des
 }
and, in general, that any two polynomials of \m { \qD } commute, the
obtained sum can be written as
 \eq{.17.71.}{
\0 1 { \Dss \Ded + 2 \Dsd \Des } \qsig\para = 3 \Ded \Des \qeps\para .
 }

In the example of \re{.17.66.}--\re{.17.67.}, \re{.17.71.} reads, after
normalizing the coefficient of \m { \qsig\para } to 1,
 \eqan{@206.}{
\qsig\para +
\f{\qbet\dev + 2 \qbet\sph + \qtau\sph \qalp\dev + 2 \qtau\dev
\qalp\sph} {\qalp\dev + 2 \qalp\sph} \dot\qsig\para
 & \mathrel +
 \leln{@207.}
\mathop +
\f{\qgam\dev + 2 \qgam\sph + \qtau\sph \qbet\dev + 2 \qtau\dev
\qbet\sph} {\qalp\dev + 2 \qalp\sph} \ddot\qsig\para
 & \mathrel +
 \lel{.18.72.}
 \mathop +
\f{\qtau\sph \qgam\dev + 2 \qtau\dev \qgam\sph}{\qalp\dev + 2 \qalp\sph}
\dddot\qsig\para
 \s=
\f{3 \qalp\sph \qalp\dev}{\qalp\dev + 2 \qalp\sph} \qeps\para +
\f{3 \0 1{\qalp\sph \qbet\dev + \qalp\dev \qbet\sph}}{\qalp\dev + 2
\qalp\sph} \dot\qeps\para +
 \leln{@209.}
 & \mathrel +
\f{3 \0 1{\qalp\dev \qgam\sph + \qbet\sph \qbet\dev + \qalp\sph
\qgam\dev}} {\qalp\dev + 2 \qalp\sph} \ddot\qeps\para +
 \leln{@210.}
  & \mathrel +
\f{ 3 \0 1{\qbet\dev \qgam\sph + \qbet\sph \qgam\dev}}{\qalp\dev + 2
\qalp\sph} \dddot\qeps{}\para + \f{3 \qgam\sph \qgam\dev}{\qalp\dev + 2
\qalp\sph} \ddddot\qeps{}\para \,.
 }
Apparently, the emerging uniaxial rheology is a \model{0, 1, 2, 3}{0,
1, 2, 3, 4} model.

As a simple but remarkable special case, if we have a \model{0}{0, 1}
Kelvin--Voigt model deviatorically and a \model{0}{0} Hooke one
spherically, the uniaxial rheology proves to be a Poynting--Thomson--Zener
model, exhibiting creep and relaxation. Consequently, \textit{deviatoric
creep is enough to produce uniaxial relaxation.}

Naturally, it is also possible to eliminate \m { \qeps\para }, in an
analogous way, to obtain a formula containing \m { \qsig\para } and its
derivatives and \m { \qeps\orth } and its derivatives. Or, \m {
\qsig\para } can be eliminated in favor of \m { \qeps\para }, \m {
\qeps\orth } and derivatives. The former formula can useful in a stress
governed process to show how transversal strain behaves, and the latter
to see how much more intricate the relationship between \m { \qeps\orth}
and \m { \qeps\para } is than what the Hookean expectation of a constant
Poisson ratio would suggest.

The technique used here to derive the uniaxial resultant rheology is
similar to how serial and parallel arrangements of rheological models
can be calculated \cite{Ful08c}. In fact, this is more than a
simple similarity: the above steps are \textit{exactly} those how one
can determine the rheology of the arrangement
 \eq{.18.73.}{
{ \1 1{\qqD \serconn \qqD \serconn \qqS} \parconn \1 1{\qqD \serconn
\qqD \serconn \qqS} \parconn \1 1{\qqD \serconn \qqD \serconn \qqS} } ,
 }
where \m { \qqD } stands for the deviatoric model \re{.17.62.}, \m { \qqS
} for the spherical one \re{.17.65.}, and \mm { \serconn } denotes the
serial and \m { \parconn } the parallel connection. The uniaxial
rheology is thus a certain serial-plus-parallel combination of the
deviatoric and the spherical rheology.

\sect{.18..B.}{Shifting internal energy}

Here, another approach to introduce an internal variable for rheology is
shown, which approach leads to the same \model{0, 1}{0, 1, 2} model,
after elimination, for constant temperature processes. For simplicity,
we exploit only the one dimensional version.

In this version, we shift internal energy, rather than entropy, with a
quadratic term \1 1 {see also \cite{Klu62a3,Klu63a,KluCia78a}} of an
internal variable \m { \qeta }, which term preserves the convexity
property:
 \eq{.19.74.}{
\qext{\qe} = \qini\qe + \f {1}{2} \qeta^2 ,
 }
which, in terms of the canonical variables, means the shifting of the
internal energy variable in any constitutive function like entropy:
 \eq{.19.75.}{
\qext{\qs} \0 1 {\qext\qe, \qeps, \qeta} = \qini\qs \0 1 {\qext\qe -
\f{1}{2} \qeta^2, \qeps} .
 }
As before \1 2 {see Eq. \re{.13.53.}}, the introduction of the  internal
variable is assumed to be accompanied by a new stress term. Taking the
time derivative of this \m { \qext\qs }, the entropy production gets
shifted now by
 \eq{.19.76.}{
\f {1}{\qT} \0 1 { \qirr\qsig \dot{\qeps} - \qrho \qeta \dot{\qeta} } .
}
Consequently, apart from a factor \m { \qT }, the Onsagerian solution
happens similarly as done previously. The constant-\m { \qT }
elimination of \m { \qeta } also leads to the linear \model{0, 1}{0, 1,
2} rheology, with the same conditions on the coefficients.

What is remarkably different in the two approaches is the balance of
internal energy, since now there is an additional, potential-like,
internal energy term \m { \f {1}{2} \qeta^2 } so a part of the
mechanical power (and of the incoming heat flux) now changes this extra
term and only the remaining part changes the specific heat related
original internal energy term \m { \qth\qe } and the elastic energy term
\m { \qel\qe }. Correspondingly, temperature changes differently than in
the entropy shifting approach. This means an experimental possibility to
distinguish between the two scenarios.

At last, it is possible to combine the two methodologies: we can permit
a quadratic shift of both entropy and internal energy. Then an extra
coefficient must be allowed in at least one of the two quadratic
expressions: the general case can be written as
 \eq{.19.77.}{
\qext{\qe} = \qini\qe + \f {\qa}{2} \qxi^2 ,
 \qquad
\qext{\qs} \0 1 {\qext\qe, \qeps, \qxi} = \qini\qs \0 1 {\qext\qe -
\f {\qa}{2} \qxi^2, \qeps} + \f {\qb}{2} \qxi^2 .
 }
The entropy production still gets shifted by a term proportional to \m
{- \qxi \dot\qxi }, now with a coefficient that contains both \m { \qa }
and \m { \qb }. The Onsagerian solution also has the same form as in the
two previous cases, and the elimination also goes through analogously.
The coefficients \m { \qa }, \m { \qb } are distinguished by their role
in the balance of internal energy, hence, in the rate equation for
temperature. This can be the basis for determining them experimentally.


\end{document}